\documentclass[a4paper,11pt]{article}
\pdfoutput=1 

\usepackage{jcappub} 

\usepackage[T1]{fontenc} 
\usepackage{xcolor}
\usepackage{ulem}

\title{\boldmath Exact solutions for compact stars with CFL quark matter}

\author[a]{L. S. Rocha,\note{Corresponding author.}}
\author[a]{A. Bernardo,}
\author[b,c]{M. G. B. de Avellar,}
\author[a]{J. E. Horvath}

\affiliation[a]{Universidade de S\~ao Paulo,\\Rua do Mat\~ao, 1226 - Butant\~a, S\~ao Paulo - SP, 03178-200, Brazil}
\affiliation[b]{Instituto Tecnol\'ogico de Aeron\'autica,\\Pra\c ca Marechal Eduardo Gomes, 50 - Vila das Ac\'acias, S\~ao Jos\'e dos Campos - SP, 12228-900, Brazil}
\affiliation[c]{Universidade Federal de S\~ao Paulo,\\Unidade Jos\'e Alencar - Rua S\~ao Nicolau, 210 - Centro - 09913-030 - Diadema - SP, Brazil}

\emailAdd{livia.silva.rocha@usp.br}
\emailAdd{antonio.bernardo@usp.br}
\emailAdd{marcio.avellar@usp.br}
\emailAdd{foton@iag.usp.br}

\abstract{The search for the true ground state of the dense matter remains open since Bodmer, Terazawa and others raised the possibility of stable quarks, boosted by Witten's \textit{strange matter} hypothesis in 1984. Within this proposal, the strange matter is assumed to be composed of $strange$ quarks in addition to the usual $up$s and $down$s, having an energy per baryon lower than the strangeless counterpart, and even lower than that of nuclear matter. In this sense, neutron stars should actually be strange stars. Later work showed that a paired, symmetric in flavor, color-flavor locked (CFL) state would be preferred to the one without any pairing for a wide range of the parameters (gap $\Delta$, strange quark mass $m_s$, and bag constant B). We use an approximate, yet very accurate, CFL equation of state (EoS) that generalizes the MIT bag model to obtain two families of exact solutions for the static Einstein field equations constructing families of anisotropic compact relativistic objects. In this fashion, we  provide exact useful solutions directly connected with microphysics.}

\begin{document}
\maketitle
\flushbottom

\section{Introduction}\label{sec:intro}
\qquad
It took just a year since Einstein's General Relativity Theory was published for Karl Schwarzschild obtain the first exact solution of Einstein's field equations. The number of valid, exact solutions has been growing since then. Different models can be constructed for a variety of applications. In cosmology, for example, solutions of a spherically symmetric spacetime have been used to model the behaviour and evolution of the early universe \cite{a}. Electromagnetic fields contributions have also been studied in solutions of Einstein-Maxwell equations and the list is still growing.

The specific set of solutions useful for the modeling of compact stars has a long and rich history. Following the discovery of the neutron \cite{Chadwick}, Landau's theoretical insight \cite{Landau} and Baade and Zwicky observational intuition \cite{BZ} started the idea of {\it neutron stars} as relativistic remnant stars of very dense matter. In 1939, Richard Tolman developed a method for treating Einstein Field equations and found new solutions for static sphere of fluids in terms of known analytic functions \cite{tolman01}, some of them of direct application to the neutron star problem. 

Tolman's kind of solutions have a highly predictive power to understand the properties of compact stars analytically. However, because of the number of degrees of freedom of the system of equations and the lack of knowledge about the microphysics of matter, his solutions do not describe properly, in modern terms, neutron stars (although some basic physical constrains were satisfied).

The problem of acceptability of the exact solutions as physical (rather than mathematical) models of real compact stars was revisited by Delgaty and Lake, who constructed a catalog containing all analytic solutions that describe isolated\footnote{The boundary occurs at a finite radius.} static spherically symmetric perfect fluid that satisfies all the necessary conditions to be physically interesting\footnote{These conditions will be discussed later in this work} \cite{b}. They concluded that out of 127 solutions, only 16 of them satisfy these conditions, and of these only 9 present a monotonic decrease of sound speed, a highly desirable if not mandatory condition.

Exact solutions are used in astrophysics to study physical features of relativistic spheres but, in spite of the consistency of this small number of relevant solutions describing a specific object, they do not provide information about their inner nature either. The main reason is that they can be rarely related to a realistic physical theory of the microphysical composition.

Generally speaking, the issue of the equation of state of matter above the nuclear saturation density, appropriate for the construction of models of compact stars, is a vast arena for discussions and calculations. However, a careful scrutiny of the proposed equations of state reveals a class of models which may be physically relevant and also allow an analytical integration of the static Einstein field equations. We shall present in this work a new class of exact solutions connected with microphysics, providing a realistic equation of state which describes the internal structure of a compact star made of strange matter in the \textit{color flavor locked} (CFL) phase.

The mass-radius M-R relation (one of the most important features of stellar sequences) will be constructed analytically, e.g. without a numerical integration of \textit{TOV} equations, and the maximum mass and its correspondent radius, a hot topic in compact star astrophysics, determined using the exact solution obtained here. Needless to say, all stellar features relevant for observational purposes can be derived from the presented expressions as well.  

\section{Color-flavor locked strange matter}
\qquad
Edward Witten \cite{c} in 1984 was one of the first, following the work of Bodmer \cite{Bod}, Terazawa \cite{Tera} and others, to boost the proposal that the strange matter coould be the true ground state of hadrons, instead of $^{56}Fe$, having a lower energy per baryon than ordinary nuclei. This matter is assumed to be composed of roughly equal numbers of \textit{up}, \textit{down} and \textit{strange} quarks, and a small number of electrons to attain the charge neutrality. If this hypothesis is correct, \textit{neutron stars} (NS) would actually be \textit{strange stars} (SS), or at least hybrid stars with a thin crust of nuclei, whenever pressure conditions in the inner layers are extreme enough to convert hadronic matter into this new stable phase of quarks. 

A great effort has been made to describe correctly and accurately the physical features of these compact objects since Witten's idea. Alcock et al. \cite{d} (hereafter referred to as AFO) extensively described an SS using the MIT bag model with a linear equation of state that does not include the strange quark mass, and that assumes quarks as asymptotically free particles. A full survey of the original SS models was published by Benvenuto and Horvath \cite{BHmnras1989}.  More recent models try to implement different contributions to the equation of state, by noting that paired states should be relevant in the dense phase. The presence of gaps actually {\it enhance} the possible stability of the quark matter phase, as discussed by Lugones and Horvath \cite{e,f} because the Fermi energy of the system is lowered by the formation of Cooper pairs. A parametric study of this possibility has been presented by Alford and collaborators \cite{MarkandCo}. 

\subsection{CFL equation of state}
\qquad
Rajagopal and Wilczek \cite{g} showed that strange matter in the CFL phase requires an equal number of \textit{u},\textit{d},\textit{s} quarks and is electrically neutral even in the presence of an electron chemical potential. In the simplest approach, the EoS is obtained by starting with the thermodynamic potential, constructed in detail in references \cite{g, h}.

The potential is given by the sum of a free quark matter potential, where quarks are assumed to have a common Fermi momentum, with a term due to the pairing interactions and a term due to the vacuum energy $B$:
\begin{equation}
\Omega_{CFL} = \frac{6}{\pi^2} \int_0^\nu p^2 \left(p-\mu \right) dp + \frac{3}{\pi^2} \int_0^\nu p^2 \left( \sqrt{p^2+m_s^2}-\mu \right) dp - \frac{3\Delta^2\mu^2}{\pi^2}+B,
\end{equation}
where the number densities are $n_u = n_d = n_s = (\nu^3 + 2\Delta^2\mu)/\pi^2$, the chemical potential is $\mu = (\mu_u + \mu_d + \mu_s)/3$ and the Fermi momentum is:
\begin{equation}
    \nu = 2\mu - \sqrt{\mu^2 + \frac{m_s^2}{3}}.
\end{equation}
Quarks \textit{u} and \textit{d} are assumed to be massless in comparison to $m_s$, which is the mass of quark \textit{s}. The gap term $\Delta$ stands for quark interactions, and we are not interested in their exact (complicated) nature.

Using this expression for $\Omega_{CFL}$, an approximation to the order $m_s^2$ of the equation of state was given in reference \cite{e}, knowing that in a degenerate state $P = - \Omega_{CFL}$ and $\rho = \sum_i \mu_i n_i + \Omega_{CFL}$, leading to:
\begin{equation}
\label{eq:x}
    P = \frac{\rho}{3} - \frac{4}{3}B + \frac{2\Delta^2\mu^2}{\pi^2} - \frac{m_s^2\mu^2}{2\pi^2},
\end{equation}
given that:
\begin{equation}
    \mu^2 = -\eta + \left( \eta^2 + \frac{4\pi^2}{9}(\rho-B) \right)^{1/2} \footnote{This expression is obtained taking the root of equation $\rho =  \sum_i \mu_i n_i + \Omega_{CFL}$, which is given as equation $(9)$ in ref \cite{e}.},
\end{equation}
where $\eta$ is defined by
\begin{equation}
\label{eq:y}
    \eta = -\frac{m_s^2}{6}+\frac{2\Delta^2}{3}.
\end{equation}

\section{Exact solutions of field equations}
\label{sec:sol}
\qquad
Finding exact solutions to the Einstein field equations is one of the fundamental problems in the general theory of relativity, and have important applications especially in astrophysics and cosmology.

The interior of a star is usually modeled as a perfect fluid, what requires the pressure to be isotropic. However, theoretical advances indicate that, at least at very high densities, deviations from local isotropy may play an important role \cite{Gleiser}. The line element for an uncharged, static and spherically symmetric fluid is given by:
\begin{equation}
    ds^2 = -c^2 e^{2\nu(r)}dt^2 + e^{2\lambda(r)}dr^2 + r^2\left(d\theta^2 + \sin^2\theta d\phi^2 \right),
\end{equation}
which, assuming anisotropy, results in the system:
\begin{equation}
\label{eq:w}
    \begin{split}
\frac{8\pi G}{c^2}\rho (r) &= \frac{2\lambda'}{r}e^{-2\lambda}+\frac{1-e^{-2\lambda}}{r^2}  \,,
\\
\frac{8\pi G}{c^4}P_r(r) &= \frac{2\nu'}{r}e^{-2\lambda}-\frac{1-e^{-2\lambda (r)}}{r^2} \,,
\\
\frac{8\pi G}{c^4}P_t(r) &= e^{-2\lambda} \left[ \nu''+\nu'^2 - \nu'\lambda' + \frac{\nu' - \lambda'}{r} \right] \,,
    \end{split}
\end{equation}
where primes denote differentiation with respect to $r$. The functions $\rho$, $P_{r}$ and $P_{t}$ stand for the energy density, radial pressure and tangential pressure respectively. This is a system of highly nonlinear partial differential equations which is difficult to integrate analytically without simplifying assumptions. In our analysis we will use units where $8\pi G = 1$ and $c = 1$, and adopt the same transformations suggested by Durgapal and Bannerji \cite{i}:
\begin{equation}
\label{eq:z}
\begin{split}
    x = Cr^2 \,,
    \qquad
    Z(x) = e^{-2\lambda(r)} \,,
    \qquad
    A^2y^2(x) = e^{2\lambda(r)},
\end{split}
\end{equation}
in order to simplify the equations.

As stated by de Avellar and Horvath \cite{j}, there are at least three different strategies to solve the system above:
\begin{itemize}
    \item[1)] If functions for the pressure, energy density or one metric element are given, it is possible to find an exact or numerical solution by integration. However, there is no control over the EoS or other similar solution functions;
    \item[2)] If an EoS is given, the integration can always be performed, at least numerically, resulting in a sequence of stellar objects and their properties;
    \item[3)] If both the EoS and an additional function ($\rho(r)$ or a function of one of the metric elements) are given, a match of the overdetermined system can be done. It is also possible to employ this route without an overdetermined system if more degrees of freedom are provided (as an electric field or pressure anisotropy).
\end{itemize}

In this work, we follow the third route since our system has the necessary extra degree of freedom (given by the anisotropic pressure) to solve it analytically.

\subsection{Criteria for physical acceptability}
\label{subsec:cond}
\qquad
As stated before, to describe a relativistic object the exact solutions must satisfy certain conditions \cite{b}:
\begin{itemize}
    \item[i.] Regularity of the gravitational potential at the origin;
    \item[ii.] The radial pressure and energy density profiles must be positive definite at the origin;
    \item[iii.] The radial pressure must vanish at some finite radius;
    \item[iv.] The energy density and radial pressure profiles must decrease monotonically from the origin to the boundary;
    \item[v.] Subluminal sound speed ($v_s^2 = dP_r/d\rho < 1)$.
\end{itemize}

These conditions guarantees the usefulness of the solutions for a realistic description of stellar models, {\it provided the equation of state also represents a physical description of the interior}. If the CFL strange matter is a candidate for such a class, we can proceed to find solutions satisfying all the Delgaty-Lake requirements and the expectation of a relevant microphysical description inside.

\section{Construction of exact models}
\qquad
In section \ref{sec:sol}, we showed viable strategies to construct an exact solution for the field equations. As mentioned, we followed the {\it third} route in our study, because of the extra degree of freedom, the anisotropic pressure. For this reason, we could supply an equation of state and an \textit{ansatz} for a metric element.

Assuming a star made entirely of CFL matter, we expanded the equation \eqref{eq:x} to obtain the EoS as a function of $\rho$ only, and that depends on the free parameters B and $\eta$\footnote{$\eta$ is given by equation \eqref{eq:y} and depends upon $m_s$ and $\Delta$.}. The expansion was made to simplify the system, resulting in an equation which is easier to treat analytically , but that remains essentially identical to the {\it exact} one \cite{e}. Thus, our EoS reads:

\begin{equation}
\label{eq:t}
    P_r = \frac{1}{3}\rho + \frac{2\eta}{\pi}\rho^{1/2} - \left( \frac{3\eta^2}{\pi^2} + \frac{4}{3}B \right).
\end{equation}
If we assume $m_s \rightarrow 0$ and noninteracting quarks, the \textit{MIT bag model} equation of state is restored.

The values we used for the free parameters $m_s$, $B$ and $\Delta$ are in agreement with the ones within the so-called {\it window of stability} calculated by Lugones and Horvath \cite{e}. This window is the $m_s - B$ plane from the zero pressure point for matter to be absolutely stable, constructed for different values of the gap $\Delta$.
We now proceed to show that exact solutions exist within metric forms already present in the literature, and construct the exact models based on them.

\subsection{Thirukkanesh-Ragel ansatz}
\label{subsec:Malaver}
Using the simplifications mentioned above, together with equations \eqref{eq:z}, we assumed the same \textit{ansatz} as proposed by Thirukkanesh and Ragel and employed by Malaver \cite{k,l}, in which:
\begin{equation}
    \label{eq:yy}
    Z(x) = \left( 1-ax \right)^n,
\end{equation}
where $a$ is a real constant and $n$ an adjustable integer. The system of equations \eqref{eq:w} is then given by:
\begin{align}
    \label{eq:q}
    \frac{\rho}{C} &= \frac{1-Z}{x} -2\Dot{Z} \,,
    \\
    \label{eq:r}
    \frac{P_r}{C} &= 4Z\frac{\Dot{y}}{y} - \frac{1-Z}{x} \,,
    \\
    \frac{P_t}{C} &= 4xZ\frac{\Ddot{y}}{y} + \left( 4Z + 2x\Dot{Z} \right)\frac{\Dot{y}}{y} + \Dot{Z} \,,
\end{align}
where dots denote differentiation with respect to $x$.

The radial pressure showed in equation \eqref{eq:r} is represented by an EoS with the form $P_r = \alpha\rho + \beta\rho^{1/2} - \gamma$, where $\alpha$, $\beta$ and $\gamma$ corresponds to the coefficients of the CFL matter equation of state \eqref{eq:t}.

The mass function of a relativistic star is given by the integral:
\begin{equation}
    \label{eq:h}
    m(r) = 4\pi \int_0^r r'^2\rho(r')dr'.
\end{equation}
The mass inside a given star corresponds to $M=m(R)$, where $R$ is the radius of the star, given by the location where the pressure vanishes.

Assuming $n=1$ a first solution was constructed, which does not satisfy the criteria ii, resulting in a constant energy density and radial pressure for any point inside the star. For this reason this case is not discussed in further detail.

In the case $n=2$ an exact solution is also obtained, with the following functions of $r$:
   \begin{align}
       \rho (r) &= aC \left( 6-5aCr^2 \right) \,,
       \\
       m(r) &= 4\pi aC\left(2r^3 - aCr^5 \right) \,,
       \\
       P_r(r) &= \alpha aC \left( 6-5aCr^2 \right) + \beta \sqrt{aC \left( 6-5aCr^2 \right)} - \gamma \,,
       \\
       e^{2\lambda(r)} &=\frac{1}{1-aCr^2} \,,
       \\
       e^{2\nu(r)} &= A^2 e^{ \left[\frac{-aC \left( 1+\alpha \right) + \gamma - \sqrt{aC \left( 6-5aC r^2 \right)}}{2aC \left(aC r^2 -1 \right)} + \frac{5\sqrt{aC}\beta \arctan \left( \sqrt{6 -5aC r^2}\right) }{2aC} \right]} \left( aC r^2-1 \right)^{-\frac{1}{2}\left( 1+5\alpha \right)} \,,
       \\
       P_t(r) &= P_r(r) + \frac{1}{4}\left[ \left(\rho(r)+P_r(r)\right) \left(\frac{2m(r) + P_r(r) r^3}{r - 2m(r)}\right) + 2P_r'(r) \right] .
    \end{align}
The product $aC$ is determined from the central density, being $aC = \frac{\rho_c}{6}$.

The regularity of gravitational potential at the origin, stated in condition i, is guaranteed by the behavior of the metric functions at $r=0$:
\begin{align}
    e^{2\lambda(0)} &= 1 \,, 
    \\
    \left(  e^{2\lambda(r)} \right)^{'}_{r=0} &= 0 \,,
    \\
    e^{2\nu(0)} &= A^2 e^{ \left[\frac{aC \left( 1+\alpha \right) - \gamma + \sqrt{6aC}}{2aC} + \frac{5\sqrt{aC}\beta \arctan \left( \sqrt{6}\right) }{2aC} \right]} \left(-1 \right)^{-\frac{1}{2}\left( 1+5\alpha \right)} = cte \,,
    \\
     \left(  e^{2\nu(r)} \right)^{'}_{r=0} &= 0.
\end{align}

\begin{figure}[htbp!]
\centering 
\includegraphics[width=3.005in]{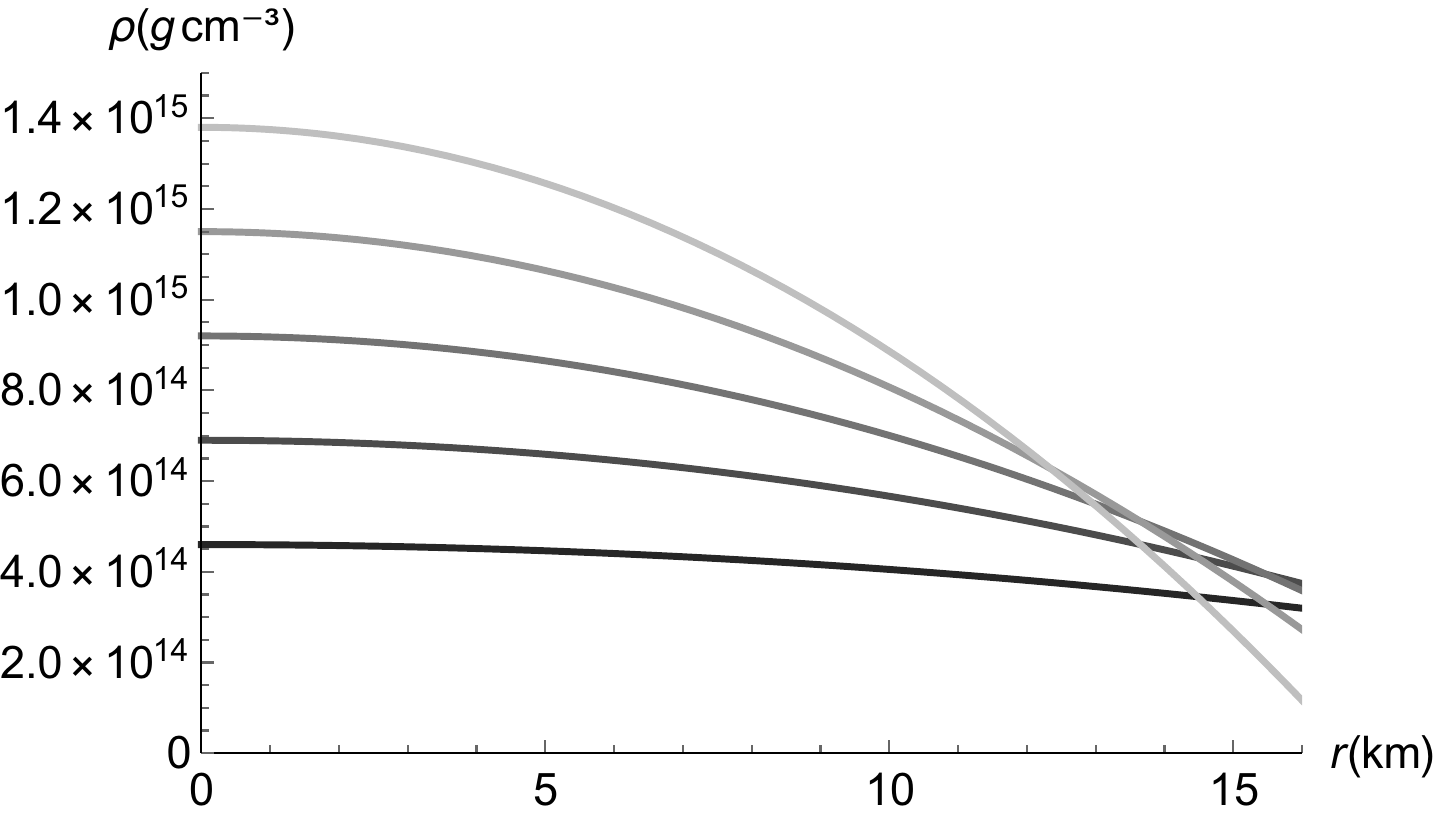}
\hfill
\includegraphics[width=3.005in]{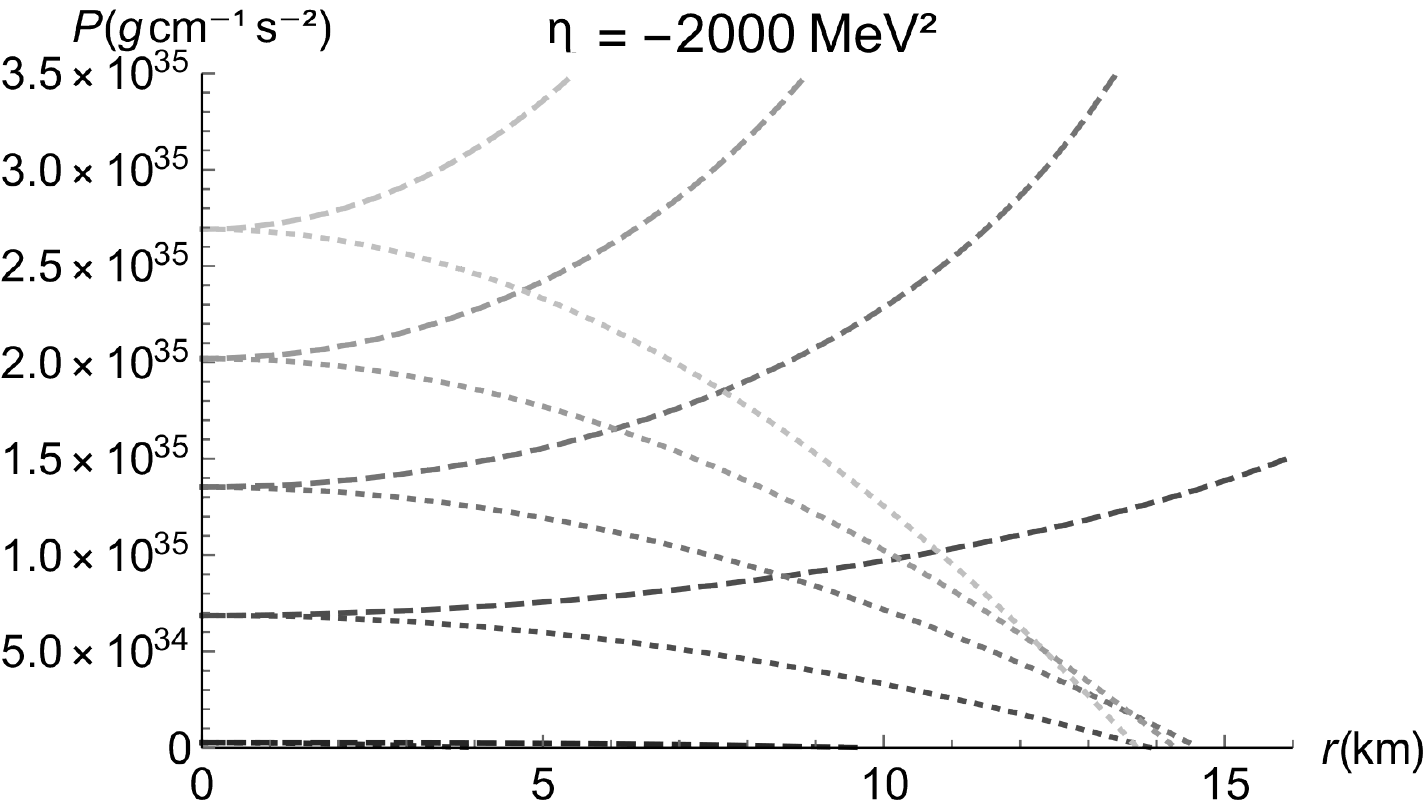}
\vfill
\includegraphics[width=3.005in]{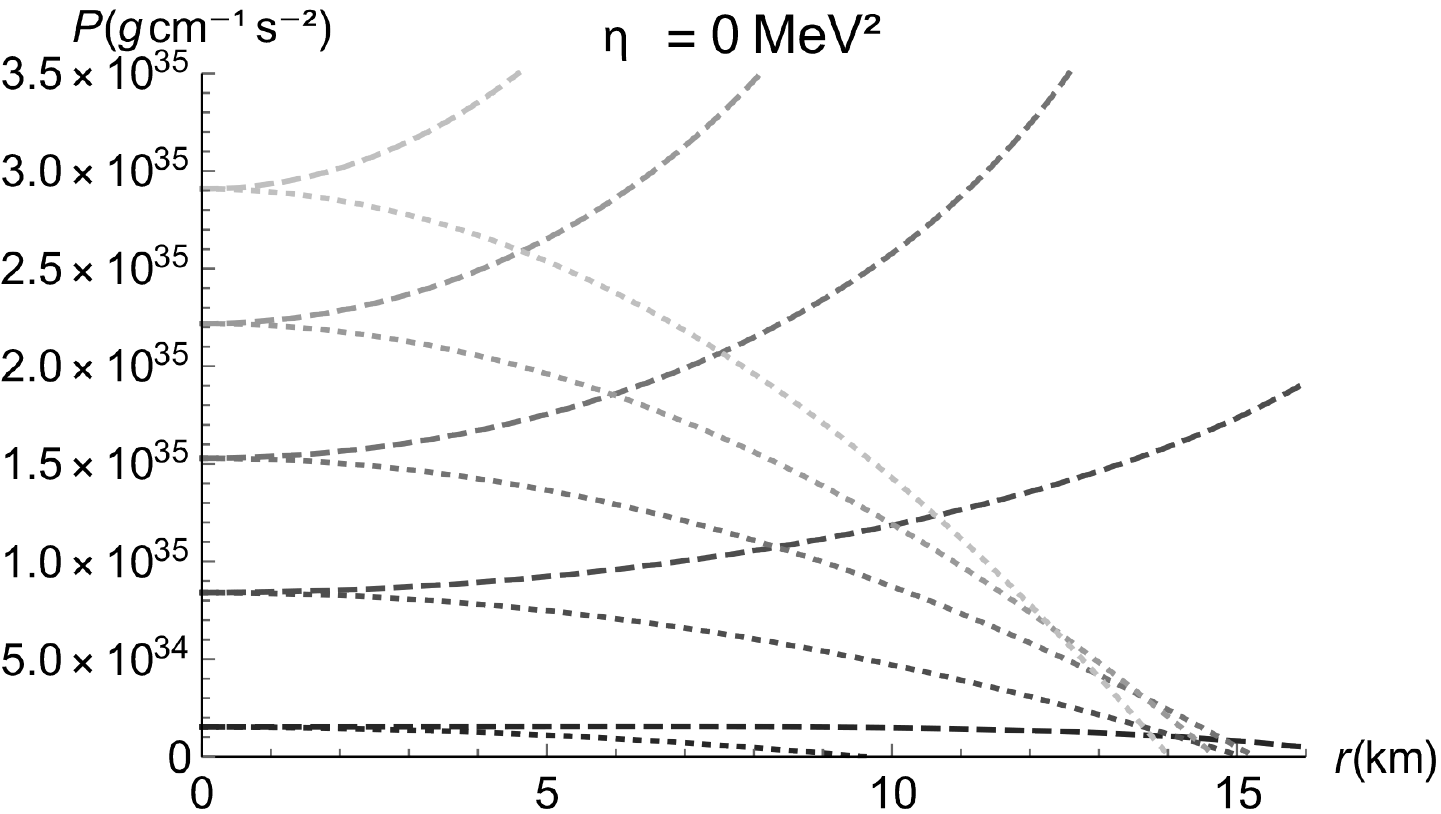}
\hfill
\includegraphics[width=3.005in]{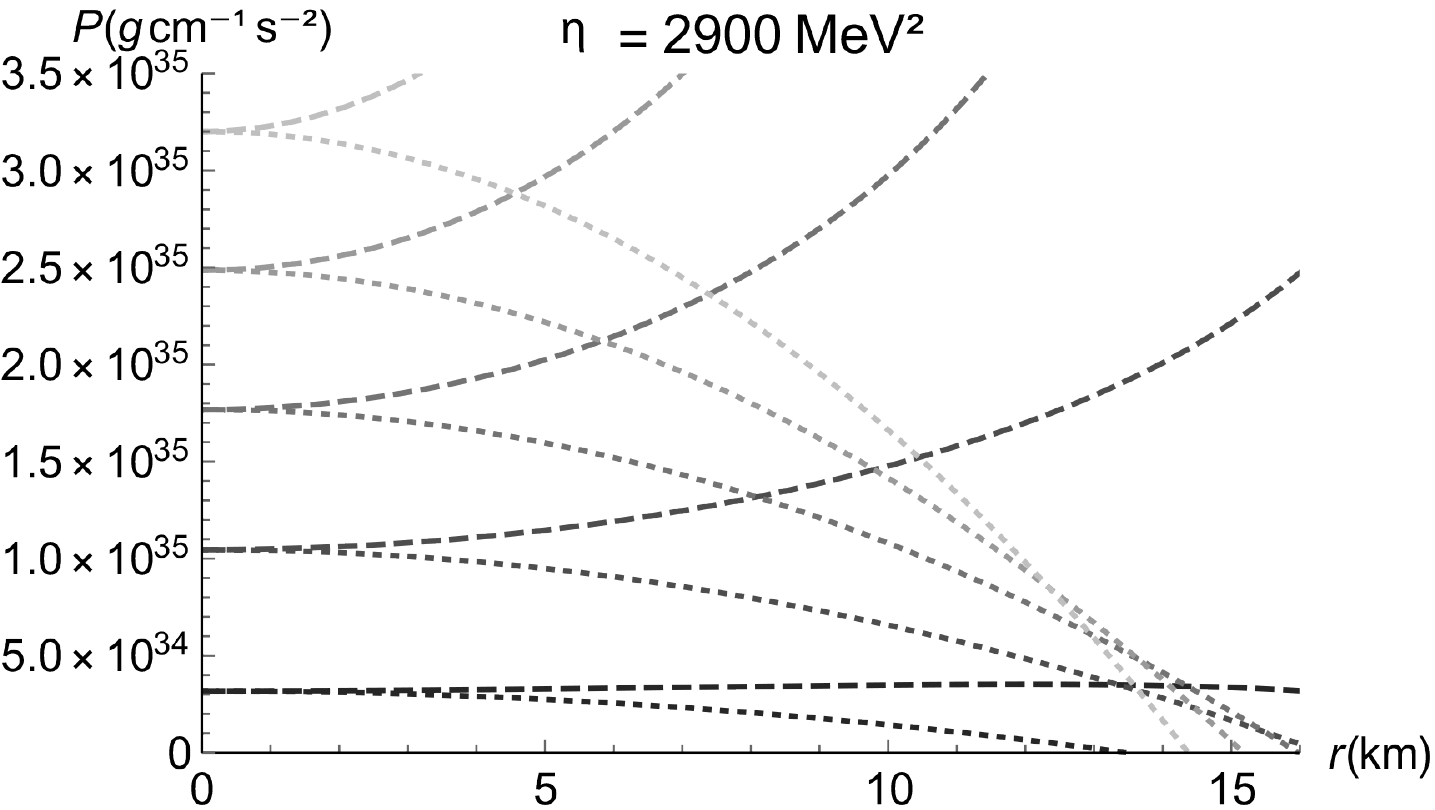}
\caption{\label{fig:i}Thirukkanesh-Ragel ansatz. The first graph (top-left) is a construction of energy density profiles monotonically decreasing with $r$, in which low central densities (darkest curves) shows a lower decrease slope. The following three graphics represents pressure profiles, being $P_r$ represented by dotted lines and $P_t$ by dashed lines. For $\eta = -2000~\mathrm{MeV^2}$ we set $m_s = 150~\mathrm{MeV}$ and $\Delta = 50~\mathrm{MeV}$; for $\eta = 0~\mathrm{MeV^2}$ we have $m_s = \Delta = 0~\mathrm{MeV}$; and finally, for $\eta = 2900~\mathrm{MeV^2}$ we fixed $m_s = 150~\mathrm{MeV}$ and $\Delta = 100~\mathrm{MeV}$. The surface of a given star is reached when $P_r$ vanish, and at this point the anisotropy factor is maximum.}
\end{figure}

Criteria ii to iv are confirmed with graphs in Figure  \ref{fig:i}, regarding that a radial pressure vanishing at a finite radius is an imposed condition to our system ($P_r(R) = 0$). We construct the energy density profile and pressure profiles for five central densities, which are set as multiples of the nuclear saturation density ($\rho_{sat} = 2.4\times 10^{14}~\mathrm{g/cm^3}$). The energy density profile (top-left panel) presents a behavior similar to what was obtained in the AFO model \cite{d} (as expected), in which for low-mass regimes $\rho$ varies very slowly with respect to $r$. When $r=R$ the energy density is finite.

\begin{figure}[htbp!]
\includegraphics[width=2.7in]{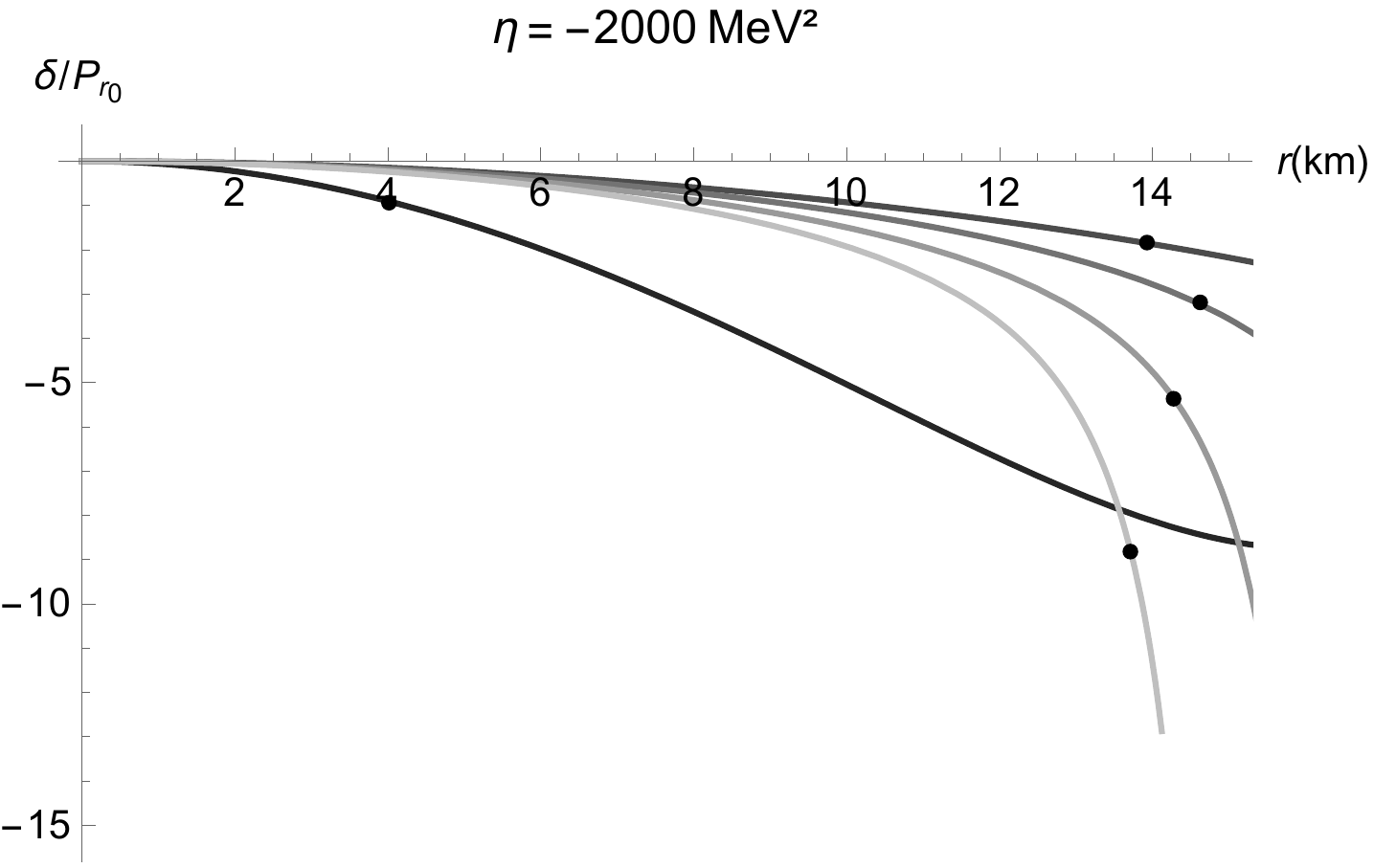}
\vfill
\centering
\includegraphics[width=2.7in]{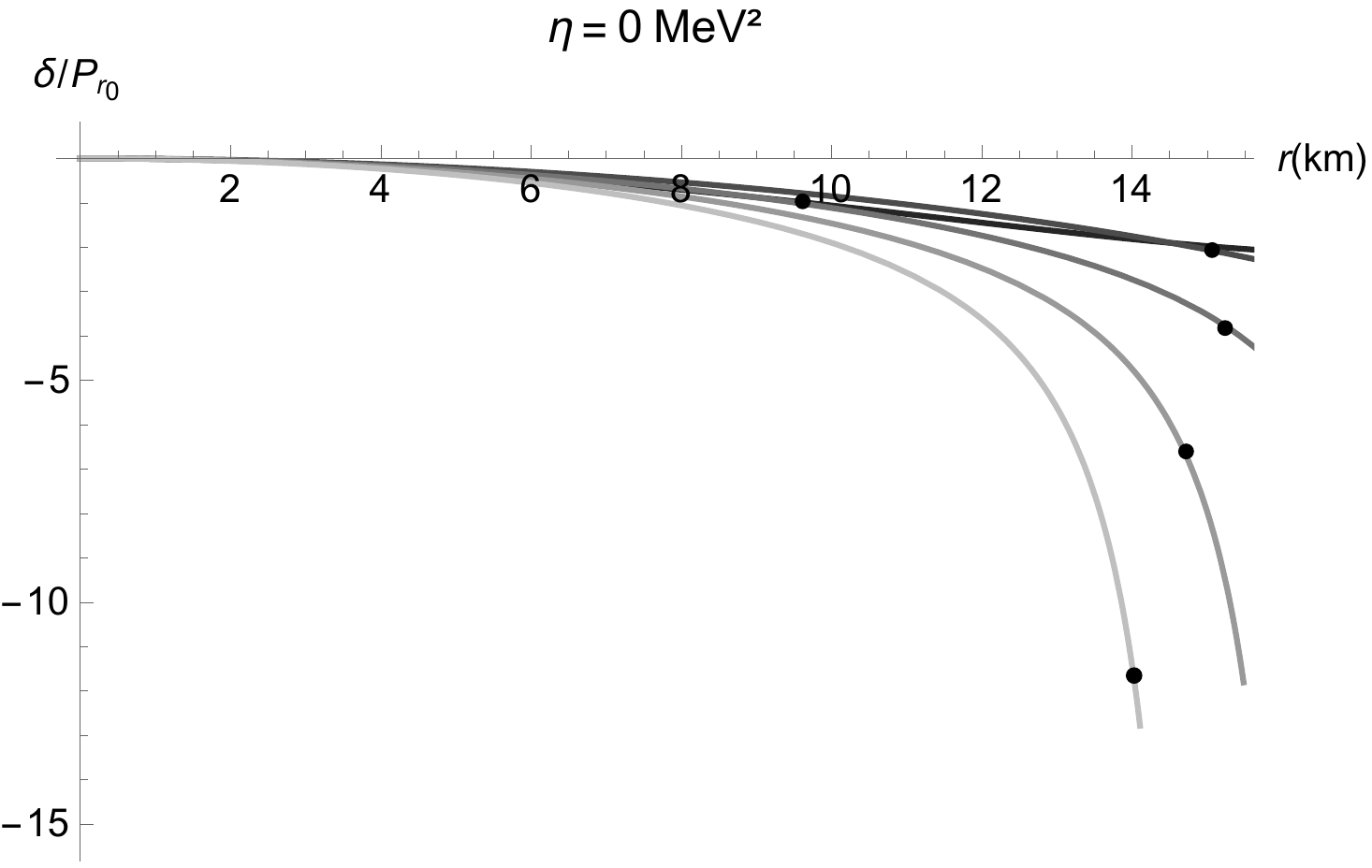}
\hfill
\includegraphics[width=2.7in]{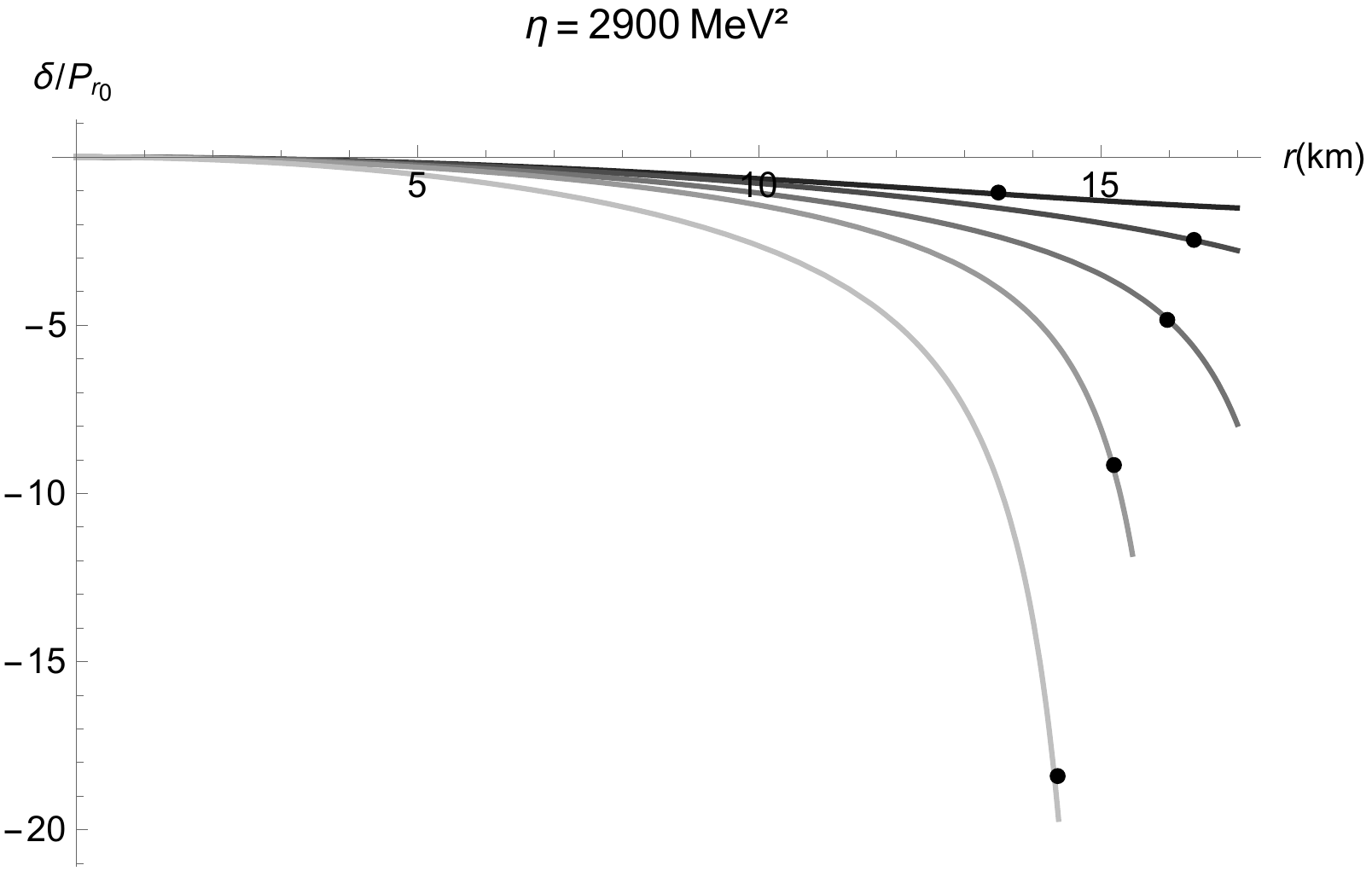}
\caption{\label{fig:TR}Thirukkanesh-Ragel ansatz. Anisotropy factor normalized by central radial pressure as a function of radial coordinate. Each graph is constructed for a given value of $\eta$, darker colors represent lower central densisties while brighter curves are for higher densities. Dots represents the point where $P_r = 0$, marking the surface of the star. As can be seen, anisotropy is maximum at the surface.}
\end{figure}

The remaining three panels of Figure \ref{fig:i} reproduce the pressure profiles, where radial and tangential pressures are represented by dotted and dashed lines, respectively. From the darkest to the brightest lines, we increased central density, starting with $2\rho_{sat}$. The additional condition discussed by Bowers and Liang \cite{m} which states that the anisotropy factor ($\delta = P_t - P_r$) must be zero for $r \rightarrow 0$, so the gradient of $P_r$ is finite in $r=0$, is also confirmed and can be seen on graphs. Moreover, it is not necessarily required that $P_t(R)$ vanishes, and like in Bowers model, in our model $\delta$ is maximum at the surface. At Figure \ref{fig:TR} we show a quantitative amount of the anisotropy factor divided by central radial pressure ($\delta /P_{r_0}$) using the same scale of grey as in Figure \ref{fig:i} to differentiate central densities. Black dots coincide with the radius of each star, meaning that $P_r = 0$ at these points. As can be noted, as high the central density, higher will be the anisotropy factor at the surface of the star, since brighter curves are related to denser objects. As expected, in this anstaz $\delta$ is very high at the surface, given that radial pressure vanishes at this point while tangential pressure reaches a maximum value. The nature of this tangential component is beyond the scope of our work but, as we will see, it has a non-negligible influence in the stellar features.
Condition v is expressed by $\alpha + \frac{\beta}{2\sqrt{\rho}} < 1$, imposing a restriction on the valid interval for $\eta$ values, being in total agreement with the range we have used in this work for a density interval between $4.5\times10^{14}$ and $2\times10^{15}~\mathrm{g/cm^3}$.

\begin{figure}[htbp!]
\centering 
\includegraphics[width=5in]{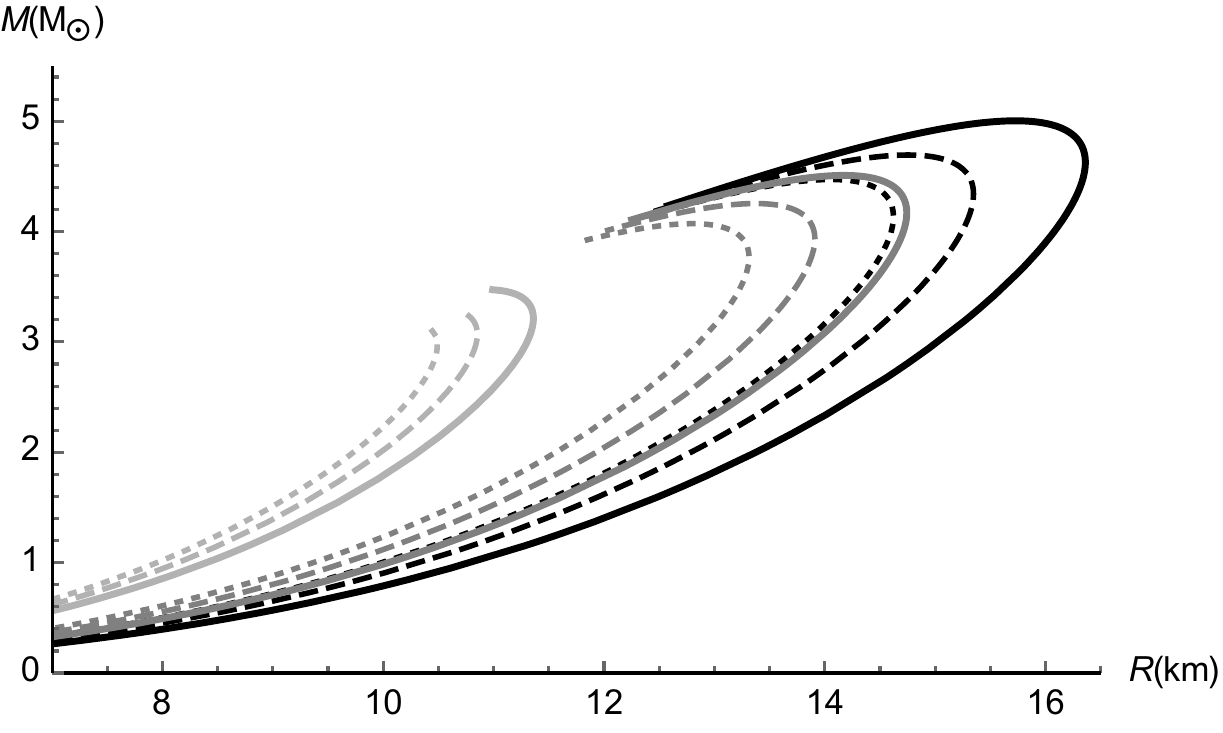}
\caption{\label{fig:k} Mass-radius relation using Thirukkanesh-Ragel ansatz. Solid lines assumes $\Delta = 100~\mathrm{MeV}$ and $m_s = 150~\mathrm{MeV}$, while the dashed lines assumes $ \Delta = m_s = 0~\mathrm{MeV}$ (resembling the MIT bag model), and for dotted lines $\Delta = 50~\mathrm{MeV}$ and $m_s = 150~\mathrm{MeV}$. From the darkest curves to the brightest we set $B = 57.5~\mathrm{MeV/fm^3}$, $B = 70~\mathrm{MeV/fm^3}$ and $B = 115~\mathrm{MeV/fm^3}$.}
\end{figure}

For this class of solutions, a mass-radius relation is constructed by finding the roots of equation \eqref{eq:t} to determine the radius of each star (given a central density), followed by integration of equation \eqref{eq:h}, and it is presented in Figure \ref{fig:k}. The M-R relation of CFL strange stars obtained for particular values of the parameters presents the same shape as the curves obtained for strange stars, a consequence of the existence of a zero pressure point at finite density. For the lowest values\footnote{Farhi and Jaffe \cite{n} concluded that the minimum value of B for deconfined matter be more stable than normal nuclear matter is $57~\mathrm{MeV/fm^3}$.} of the bag constant. The maximum mass reaches $5~\mathrm{M_{\odot}}$. Increasing the bag constant value while maintaining $\eta$ fixed has the effect of lowering both the maximum mass $M_{max}$ and the maximum radius $R_{max}$. On the other hand, an increase only in the pairing gap $\Delta$ has the effect of increasing both $M_{max}$ and $R_{max}$. Confirming the prediction obtained in \cite{f}, the combination of a large $\Delta$ with a large $B$ leads us to the most compact models of stars made of CFL strange matter. 

The behaviour of tangential pressure in this ansatz leads to high values of maximum mass (for each set of parameters), reaching values at the order of $5 M_{\odot}$, which is currently seen as prohibited for this class of stars.

\subsection{Sharma-Maharaj ansatz}
\qquad
Given the results obtained for the previous {\it ansatz}, we also tested the EoS \eqref{eq:t} on a density profile given by Sharma \cite{o} to find exact solutions:
\begin{equation}
\label{eq:m}
\rho(r) = \frac{b(3 +ar^{2})}{(1 + ar^{2})^{2}},
\end{equation}
where the parameter $b$ is defined as a function of central density, $b= \frac{\rho_{c}}{3}$, and $a$ will be a implicit function of $\rho_{c}$ carefully chosen to control the anisotropy, as we shall show later.

The exact expressions for this model are, in addition to \ref{eq:m}:
\begin{align}
      m(r) &= \frac{br^3}{2\left(1+ ar^2\right)} \,,
      \\
      P_r(r) &= \alpha\frac{b(3 +ar^{2})}{(1 + ar^{2})^{2}} + \beta \frac{\sqrt{b(3 +ar^{2})}}{\left(1 + ar^{2}\right)} - \gamma \,,
      \\
      e^{2\lambda(r)} &=\frac{1+ar^2}{1+\left(a-b\right)r^2} \,,
      \\
      e^{2\nu(r)} &= e^{Wr^2}\left( 1+ar^2\right)^{1/3}\left(1+ (a-b)r^2\right)^Y K\,,
      \\
   P_t(r) &= P_r(r) + \frac{1}{4}\left[ \left(\rho(r)+P_r(r)\right) \left(\frac{2m(r) + P_r(r) r^3}{r - 2m(r)}\right) + 2P_r'(r) \right],
   \end{align}
where $W$,$ Y$ and $K$ are functions of $\alpha$, $\beta$ and $\gamma$. We refer to \cite{o} for the actual forms of W, Y and K.

\begin{figure}[htbp!]
\centering 
\includegraphics[width=4.8in]{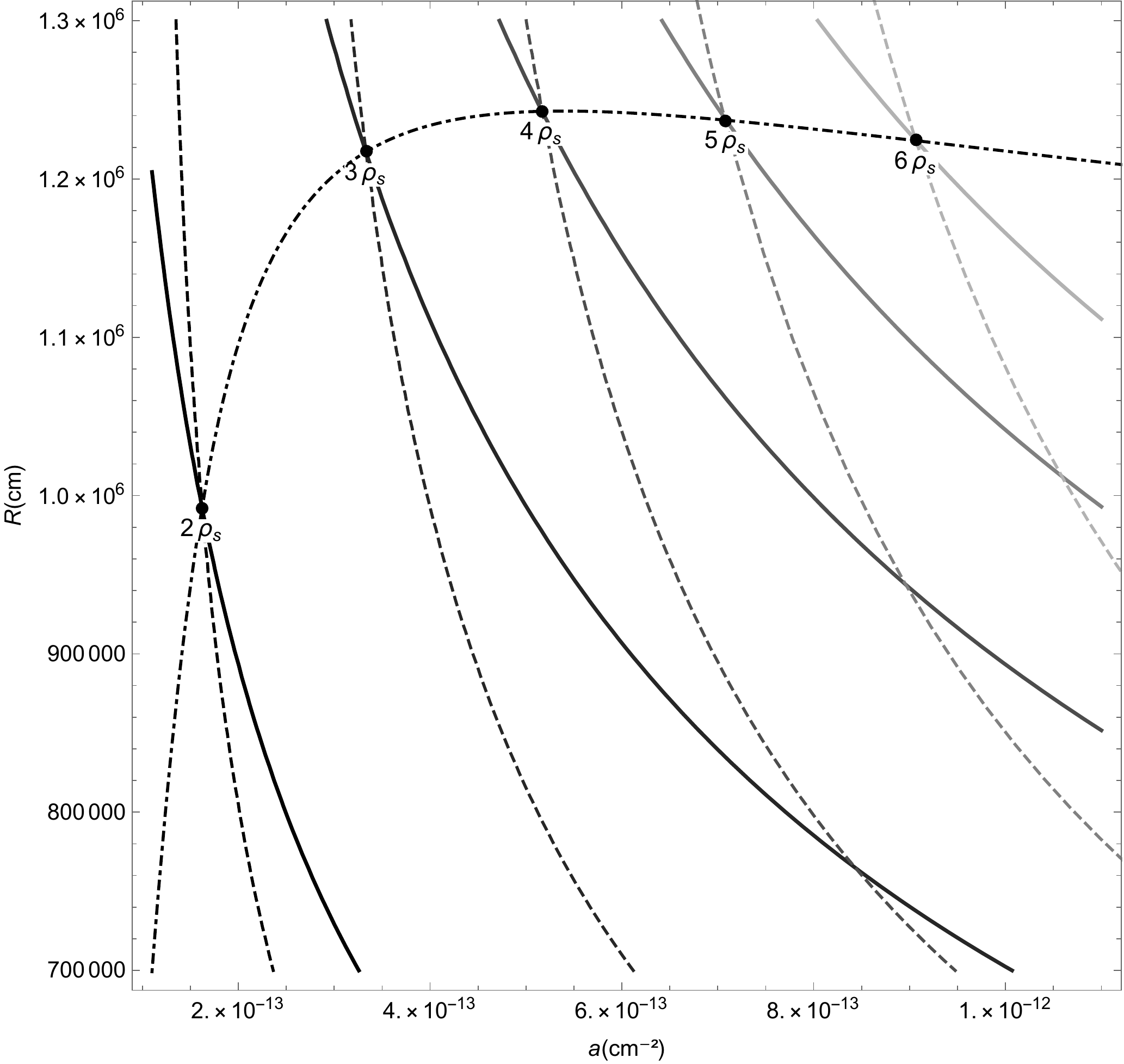}
\caption{\label{fig:j}Sharma-Maharaj ansatz. Contour plot of $P_r$ (dashed) and $P_t$ (solid) functions for five different central densities. For each $\rho_c$ there is a particular $a$ for which $P_t(R)$ vanishes. Dot-dashed curve guarantee that both pressures vanish at the surface of a star.}
\end{figure}

Panel \ref{fig:j} shows the contour plots of  $P_t$ (dashed) and  $ P_r$ (solid) for various central densities. For any central density there is a particular $a$ that vanishes $P_t(R)$. The dot-dashed line is the curve $a(\rho_c)$  that guarantees both pressures to vanish at the same point.

This curve separates panel \ref{fig:j} into two regions. If the parameter $a$ is chosen in the upper region for every  $\rho_c$, the tangential pressure will be finite at the surface of the star, similar to the model of the previous section. If, otherwise, we chose $a$ in the lower region, the tangential pressure will vanish $inside$ the star. In this work, we are not studying neither of these two regimes but we should mention that $P_{t}$ vanishing inside the star could represent the real radius of the star or the beginning of a negative tangential pressure regime close to the surface of the star. 
The models in which $P_{t}$ do vanish inside the star may still be physically acceptable, but the conditions for this to happen have not been addressed here and remain a subject for future work. A detailed discussion about stability of anisotropic objects in both regime ($P_t$ positive and negative at the surface) can be find in reference \cite{Arbanil}.

Thus, we choose the $a(\rho_c)$ that vanishes $P_t$ at the border, corresponding to the dot-dashed line. This choice yields a \textit{quasi-isotropic} model that will be the only explored henceforth.

\begin{figure}[htbp!]
\centering 
\includegraphics[width=3.005in]{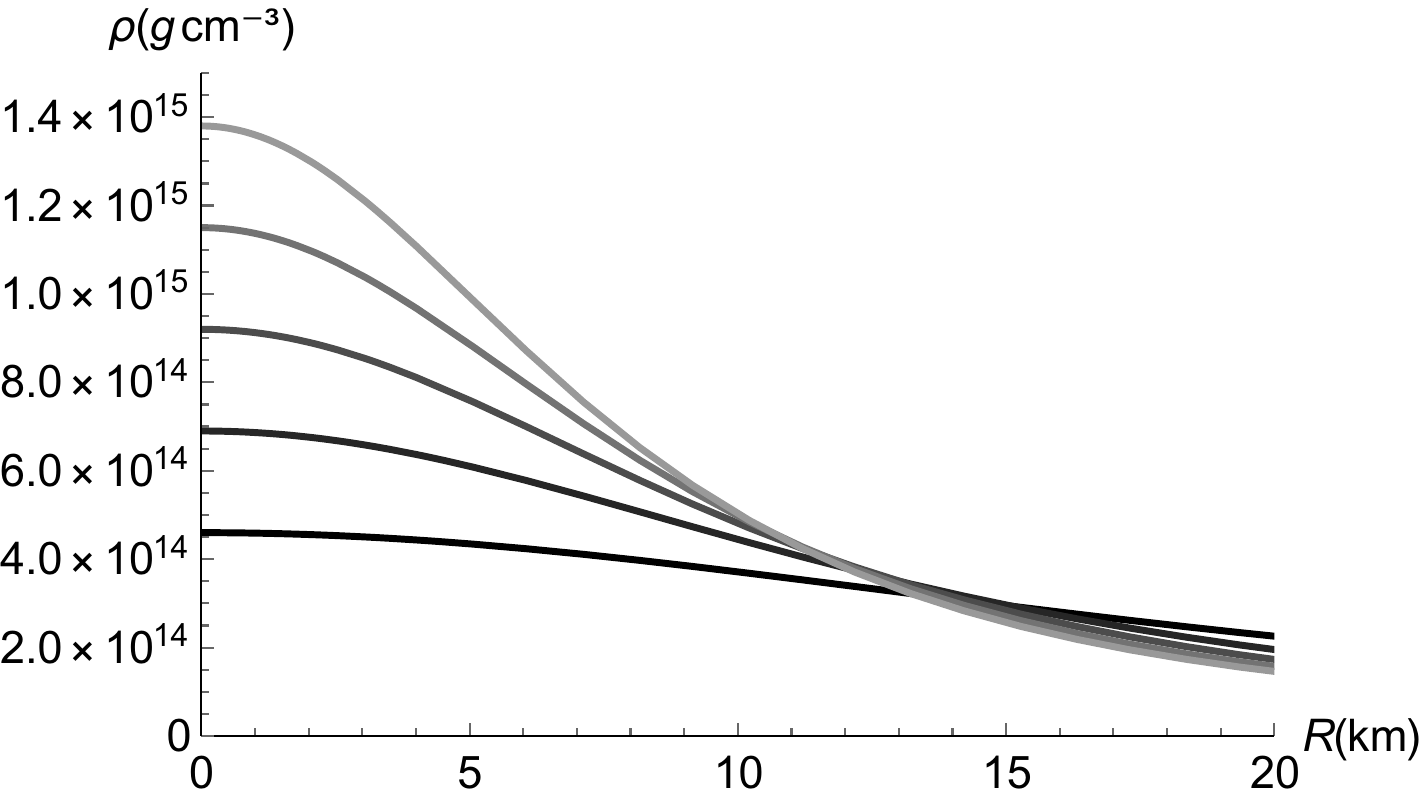}
\hfill
\includegraphics[width=3.005in]{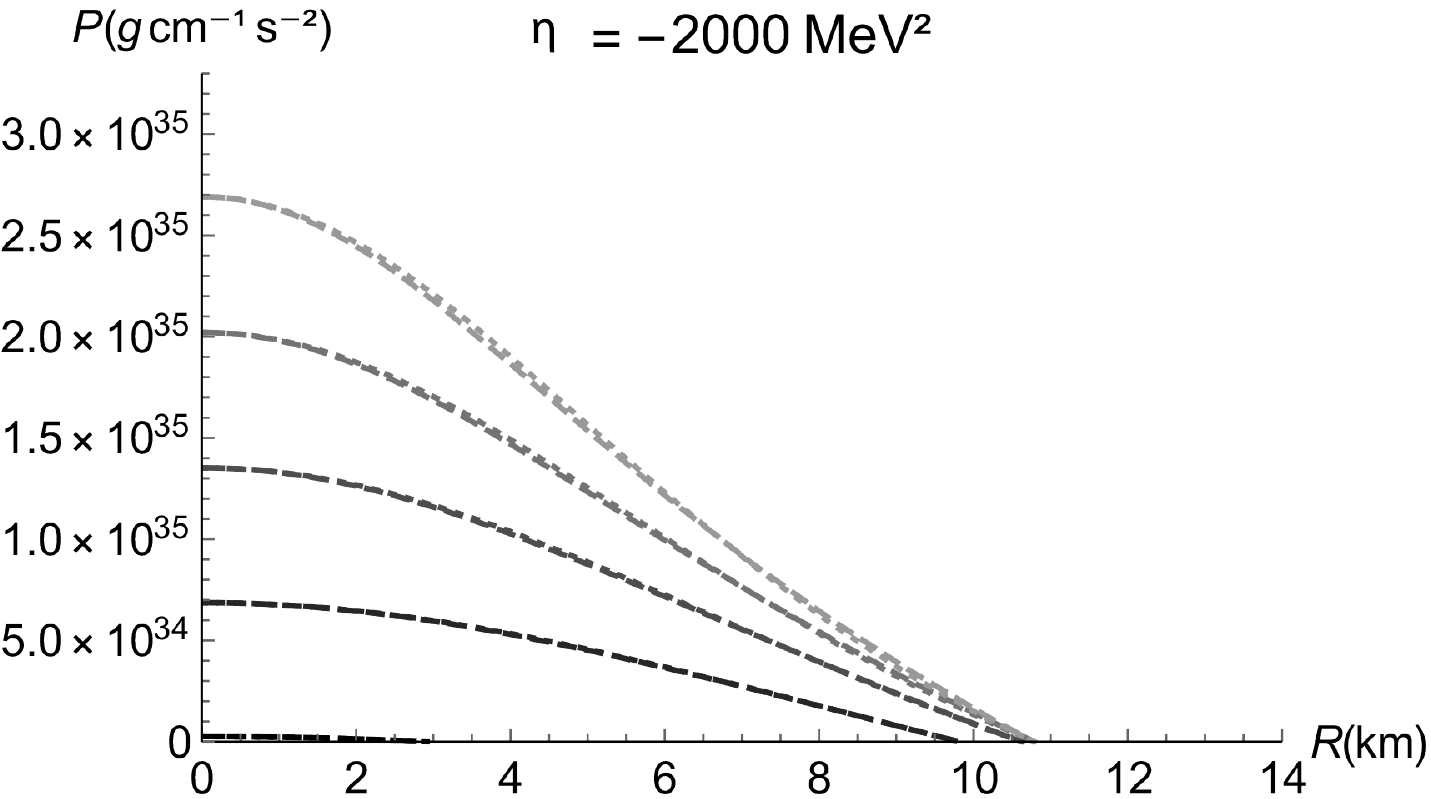}
\vfill
\includegraphics[width=3.005in]{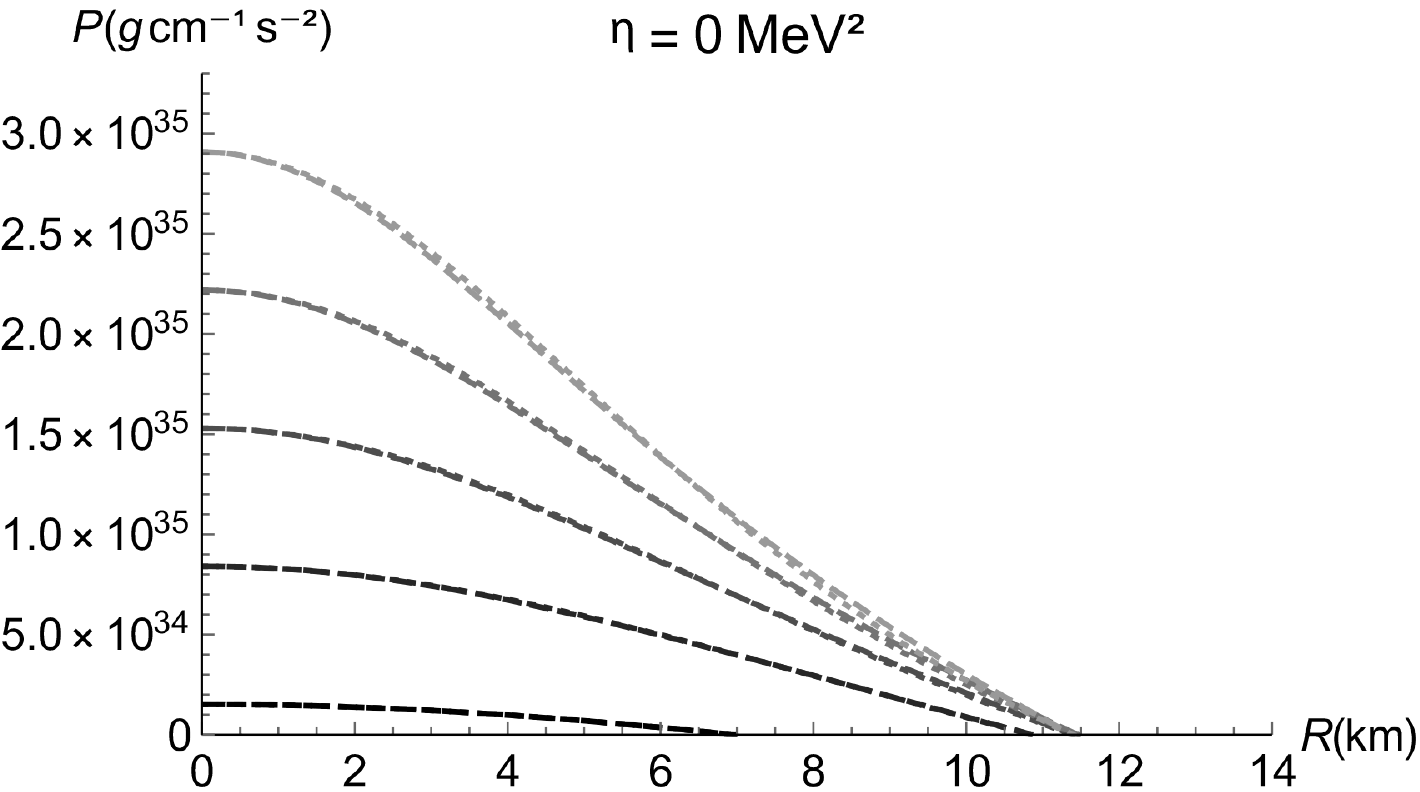}
\hfill
\includegraphics[width=3.005in]{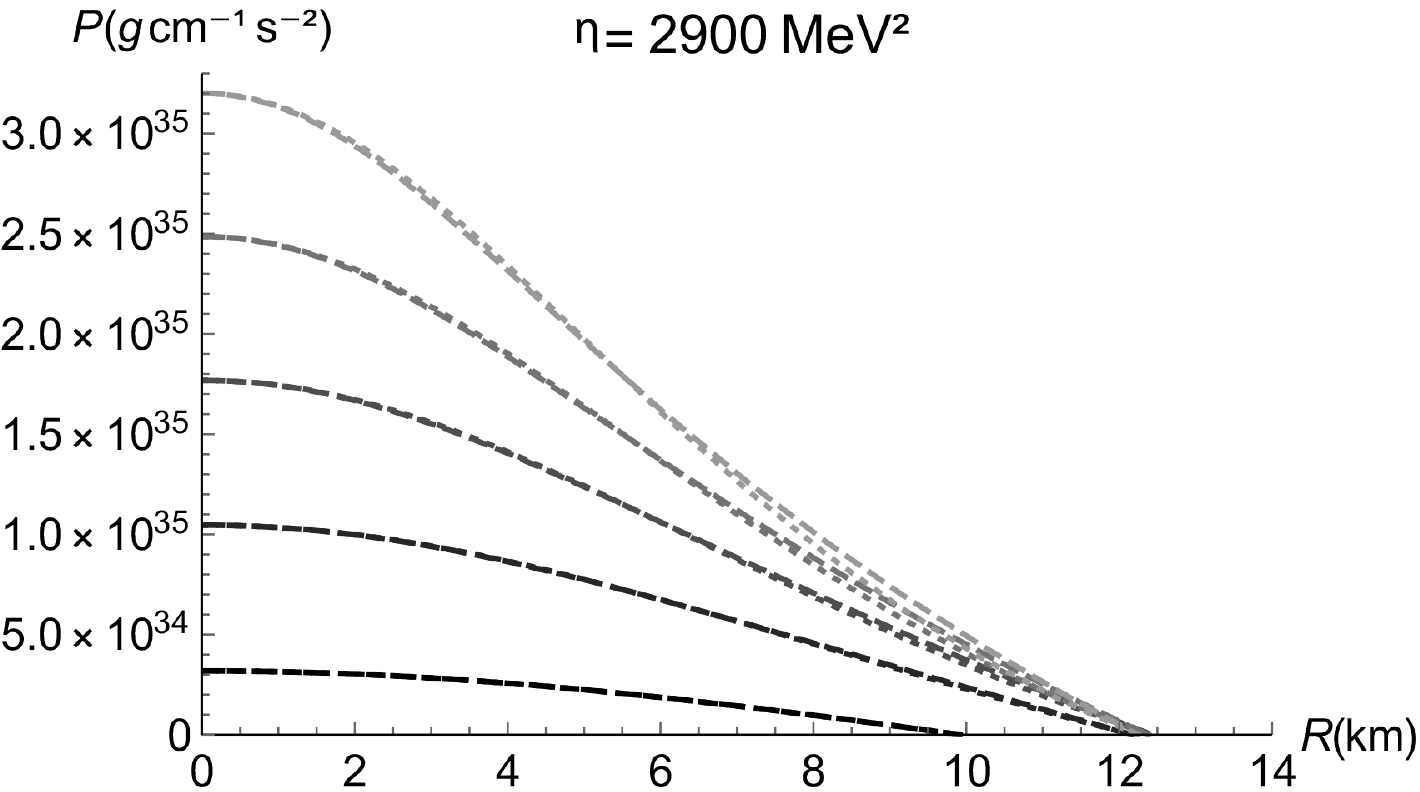}
\caption{\label{fig:l}Sharma-Maharaj ansatz. The first graph (top-left) is a construction of energy density profiles monotonically decreasing with $r$, in which low central densities (darker curves) shows a lower decrease slope. The following three graphics represents pressure profiles, with $P_r$ represented by dotted lines and $P_t$ by dashed lines. For $\eta = -2000~\mathrm{MeV^2}$ we set $m_s = 150~\mathrm{MeV}$ and $\Delta = 50~\mathrm{MeV}$; for $\eta = 0~\mathrm{MeV^2}$ we have $m_s = \Delta = 0~\mathrm{MeV}$; and finally, for $\eta = 2900~\mathrm{MeV^2}$ we fixed $m_s = 150~\mathrm{MeV}$ and $\Delta = 100~\mathrm{MeV}$. The surface of a given star is reached when $P_r$ vanishes,  at this point the anisotropy factor also vanishes.}
\end{figure}

The validity of criteria i about regularity of gravitational potential can be easily checked. Criteria ii to iv are confirmed for this particular \textit{ansatz} by the profiles showed in Figure \ref{fig:l}. Criteria v is also satisfied in the regime of free parameters that are physically acceptable.

\begin{figure}[htbp!]
\includegraphics[width=2.7in]{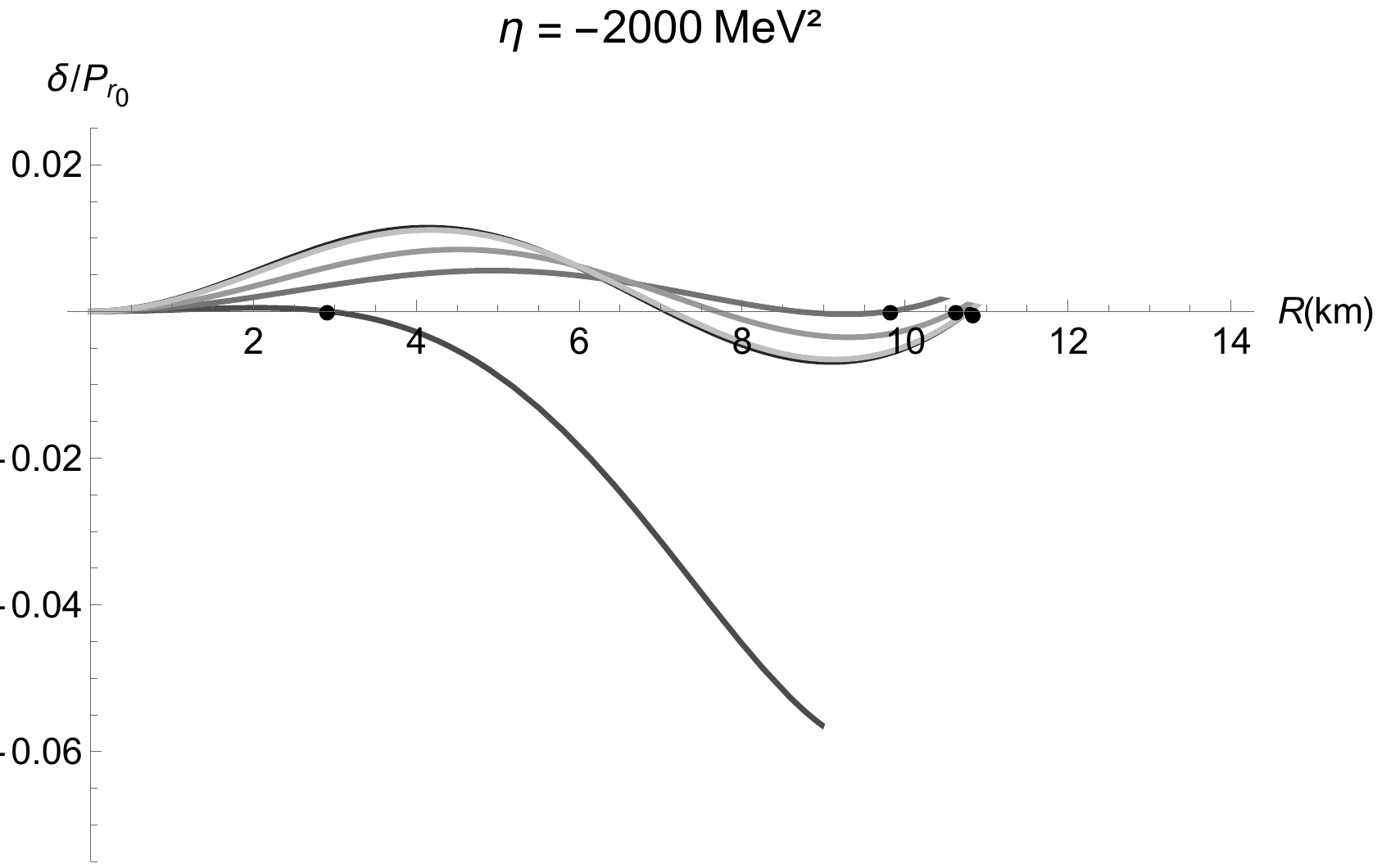}
\vfill
\centering
\includegraphics[width=2.7in]{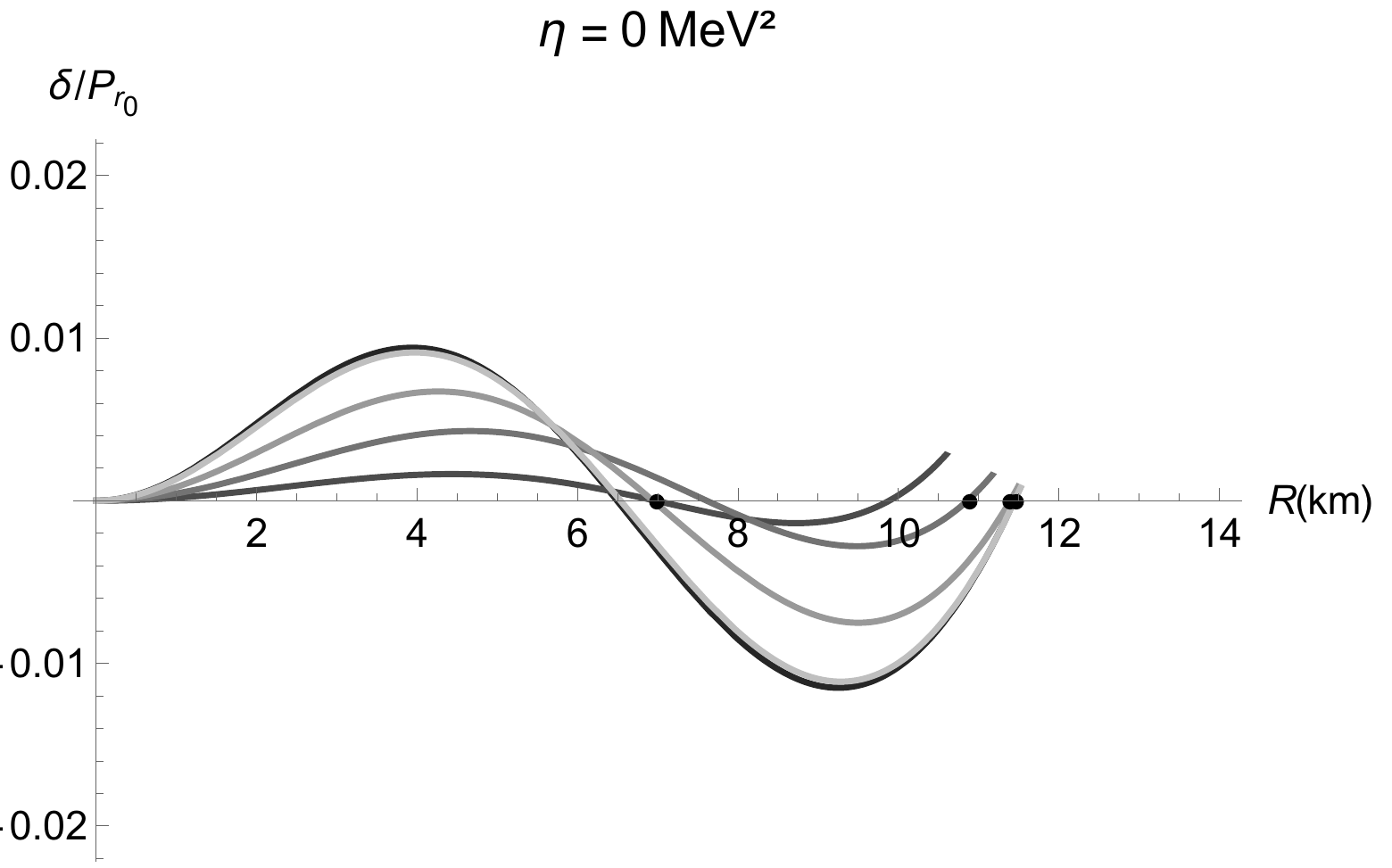}
\hfill
\includegraphics[width=2.7in]{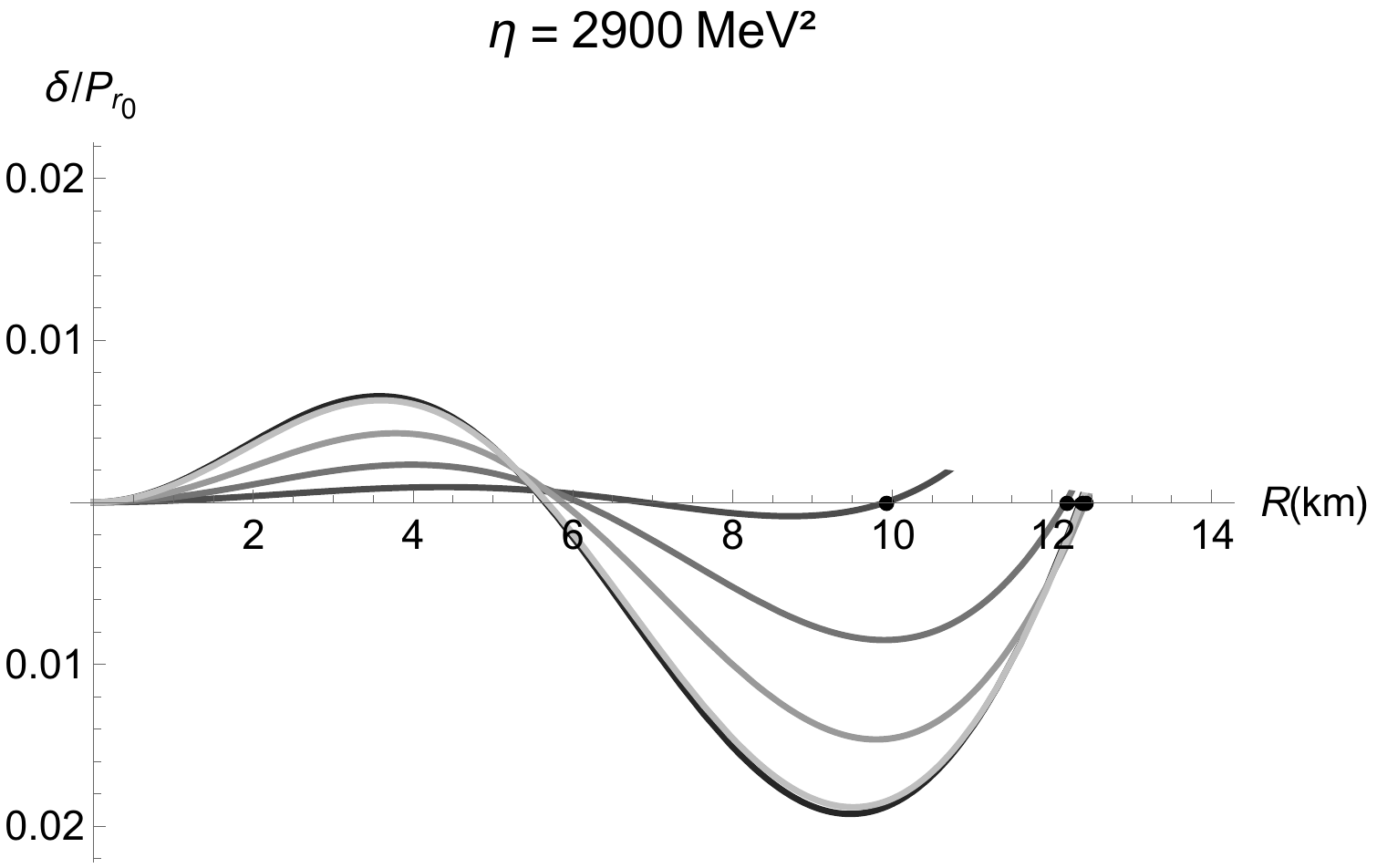}
\caption{\label{fig:SM}Sharma-Maharaj ansatz.  Anisotropy factor normalized by central radial pressure as a function of radial coordinate. Each graph is constructed for a given value of $\eta$, darker colors represent lower central densisties while brighter curves are for higher densities. Dots represents the point where $P_r = 0$, marking the surface of the star.}
\end{figure}

The energy density profiles (top-left panel of \ref{fig:l}) present the usual behaviour of being more concentrated with growing $\rho_c$. The other three panels present both the radial and the tangential pressure for different values of the parameter $\eta$ . By construction, they coincide at the border of the star. Changing the parameter $\eta$ results in a change of the radius of the star for a given $\rho_c$. It is noticeable that in this model the difference between the radial (dotted) and tangential (dashed) pressures is very mild, and a quantitative amount of the anisotropy factor is given at Figure \ref{fig:SM}, where brighter lines represent denser stars and dots mark $P_r = 0$, representing the radius of the star. Although the anisotropy factor behaviour, starting positive $(P_r > P_t)$ and turning into negative $(P_r < P_t)$ after some radius, the difference between this values is on the order of $1\%$ and therefore can be neglected.

\begin{figure}[htbp!]
\centering 
\includegraphics[width=5in]{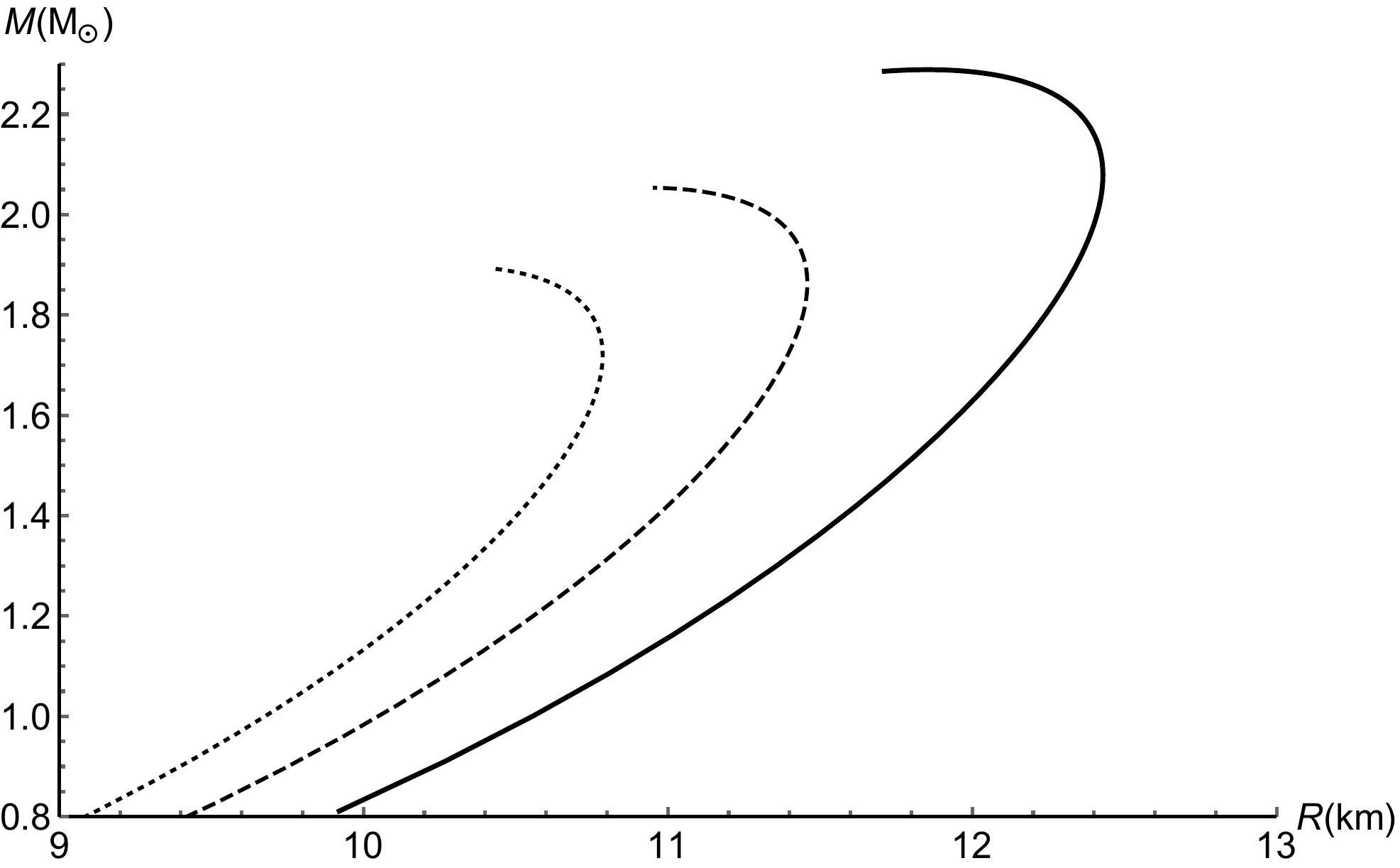}
\caption{\label{fig:o}Mass-radius relation in Sharma-Maharaj ansatz. Solid lines assumes $\Delta = 100~\mathrm{MeV}$ and $m_s = 150~\mathrm{MeV}$,  the dashed lines assumes $ \Delta = m_s = 0$ (resembling the MIT bag model), and for dotted lines $\Delta = 50~\mathrm{MeV}$ and $m_s = 150~\mathrm{MeV}$. All of the curves were done for $B = 57.5~\mathrm{MeV/fm^3}$.}
\end{figure}

Besides the fact that this model leads to an almost isotropic configuration, the most fundamental result refers to the M-R diagram (Figure \ref{fig:o}), where the maximum mass (using $B = 57.5~\mathrm{MeV/fm^3}$) is around $2.2~\mathrm{M_{\odot}}$, being in complete agreement with observed masses from pulsars\footnote{The web page https://stellarcollapse.org/nsmasses presents a compilation of observed NS masses.}. The three shown curves differ only in $\eta$ values, where the dotted one is referent to $\eta = -2000~\mathrm{MeV/fm^3}$, the dashed is for $\eta = 0~\mathrm{MeV/fm^3}$ and the solid for $\eta = 2900~\mathrm{MeV/fm^3}$. We emphasize that $\eta = 0~\mathrm{MeV/fm^3}$ resemble the MIT bag model, where strange quarks are assumed to be massless and free.

\section{Discussion}
\qquad
This work focused on the construction of exact solutions for stellar models assuming the CFL strange matter phase and allowing anisotropy in the pressure. The CFL phase at zero temperature is modeled as an electrically neutral and colorless gas of quark Cooper pairs that allows this matter to be the true ground state of strong interactions for a wide range of the parameters $B$, $m_s$ and $\Delta$. There is a large number of works, as those referenced in \cite{b}, that propose different classes of solutions for the Einstein field Equations using {\it toy models}. The novel feature of our work is to provide a realistic equation of state for the compact object, thus connecting micro and macrophysics and finding exact descriptions.

In the previous work made by Lugones and Horvath \cite{f} they have constructed a mass-radius relation for the same set of parameters, but from a numerical integration of TOV equation, what means that pressure is assumed to be isotropic from the very beginning. The maximum mass obtained in their work using $B = 70~\mathrm{MeV}$ is around $2~ \mathrm{M_{\odot}}$ while in our anisotropic model whith Thirukkanesh-Ragel {\it ansatz} can be up to $4.2~\mathrm{M_{\odot}}$. The maximum radius is also bigger in comparison with the isotropic model for all values of $B$. This effect of increasing both $M_{max}$ and $R_{max}$ is expected, as suggested by the pioneer work from Bowers and Liang \cite{m} about anisotropic models. As stated before, the nature and consequent effects of an anisotropy in pressure are still not clear and more detailed studies are necessary. 

If we compare CFL M-R relations obtained for both {\it ansatz} (Figures \ref{fig:k} and \ref{fig:o}) assuming the minimum value of $B$, the \textit{quasi-isotropic} model has a maximum mass that is about half of the maximum mass found in the anisotropic construction, for each combination of parameters $m_s$ and $\Delta$. The presence of two {\it a priori} free parameters in the Sharma-Maharaj ansatz ($a$ and $b$) compared to the presence of only one free parameter in the Thirukkanesh-Ragel model ($a$) allowed us to construct M-R relation that, regarding the observed values of pulsar masses, falls within a narrower range, at the expense of forcing both pressures to vanish at the surface, a condition that is not mandatory, but mimics the numerical results for the isotropic case.

In summary, we have provided  two anisotropic pressure models made of CFL strange matter, showing that large masses and radii are achieved. All present observational bounds on these quantities are satisfied by these models, and a number of additional requirements being evaluated (coming from X-ray burst light curves, QPOs and other phenomena actually observed) can be addressed using the exact expressions. Given the rarity of exact models, as discussed by Delgaty and Lake, we hope these can be useful for the current research on compact stars. 

\acknowledgments
\qquad
L. Rocha acknowledges the Instituto de Astronomia,
Geof\'isica e Ci\^encias Atmosf\'ericas de S\~ao Paulo and the
financial support received from Capes Agency (Brazil). A. Bernardo and J.E. Horvath acknowledges the Fundac\~ao de Amparo
\`a Pesquisa do Estado de S\~ao Paulo for partial financial support. MGBA acknowledges CNPq Project 150999/2018-6 for financial support. The authors acknowledge the FAPESP Thematic Project 2013/26258-4.



\end{document}